\documentstyle[aps,twocolumn]{revtex}

\begin{document}
\title{Scheme for preparation of mulipartite entanglement of atomic ensembles%
}
\author{Peng Xue\thanks{%
Email address: xuepeng@mail.ustc.edu.cn}, and Guang-Can Guo\thanks{%
Email address: gcguo@ustc.edu.cn}}
\address{Key Laboratory of Quantum Information, University of Science and\\
Technology\\
of China, Hefei 230026, P. R. China}
\maketitle

\begin{abstract}
\baselineskip12ptWe describe an experimental scheme of preparing
multipartite W class of maximally entangled states between many atomic
ensembles. The scheme is based on laser manipulation of atomic ensembles and
single-photon detection, and well fits the status of the current
experimental technology. In addition, we show one of the applications of the
kind of W class states, teleporting an entangled state of atomic ensembles
with unknown coefficients to more than one distant parties, either one of
which equally likely receives the transmitted state.

PACS number(s): 03.67.-a, 03.65.Ud, 42.50.Gy
\end{abstract}

\baselineskip12ptQuantum entanglement is one of the most striking features
of quantum mechanics. The recent surge of interest and progress in quantum
information theory allows one to take a more positive view of entanglement
and regard it as an essential resource for many ingenious applications such
as quantum computation \cite{Shor,Grover,Ekert1}, quantum teleportation \cite%
{Bennett1,Pan1}, superdense coding \cite{Bennett3}, and quantum cryptography 
\cite{Ekert2,xue1,xue2}. The technology of generation and manipulation of
bipartite entangled states has been realized in some systems \cite%
{Kwiat,Kwiat1,Kwiat2,other}. Recently, there has been much interest in using
quantum resource to get more and more subsystems entangled \cite%
{Bou,Rau,Sac,Pan2} for more useful applications \cite{Gottesman,Nielsen}. In
most of the above schemes, the subsystems are taken as single-particle.
Remarkably, Lukin and Duan {\it et al. }have proposed some schemes \cite%
{Lukin,Duan1,Duan2,Duan3} for preparation of entanglement which use atomic
ensembles with a large number of identical atoms as the basic system. For
example, one can use atomic ensembles for generation of substantial spin
squeezing \cite{spin} and continuous variable entanglement \cite{Duan1,Jul},
and for efficient preparation of Einstein-Podolsky-Rosen (EPR) \cite{Duan2}
and Greenberger-Horne-Zeilinger (GHZ) type of maximally entangled states 
\cite{Duan3}. The schemes have some special advantages compared with other
quantum information schemes based on the control of single particles \cite%
{Duan4}. However, there is not any scheme for experimental realization of W
class states in this system.

It is well known that there are two different kinds of genuine tripartite
entanglement-----GHZ state and W state \cite{Dur}. Indeed, any (non-trivial)
tripartite entangled state can be converted, by means of stochastic local
operations and classical communication (SLOCC), into one of two standard
forms, namely either the GHZ state or else the W state, and that this splits
the set of the genuinely trifold entangled states into two sets which are
unrelated under local operations and classical communication (LOCC). That
is, the W state can not be obtained from a GHZ state by means of LOCC and
thus one could expect, in principle that it has some interesting,
characteristic properties. The entanglement of the W class state is
maximally robust under disposal of any one of three qubits, in the sense
that the remaining reduced density matrices retain, according to several
criteria, the greatest possible amount of entanglement, compared to any
other state of three qubits, either pure or mixed. So it is important to
prepare the W class of entangled state experimentally.

In this report, we describe an experimental scheme of preparing multipartite
W class of maximal entanglement between atomic ensembles. The scheme
involves laser manipulation of atomic ensembles, beam splitters, and
single-photon detection, and well fits the status of the current
experimental technology. The first step of this scheme is to entangle two
atomic ensembles in an EPR state, which is based on the techniques proposed
in Ref. \cite{Duan2}. To prepare the W class of maximally entangled states,
two laser pulses (pumping laser and repumping laser) are applied to the
atomic ensembles and the corresponding Raman transition $\left\vert
g\right\rangle \rightarrow \left\vert s\right\rangle $ and anti-Raman
transition $\left\vert s\right\rangle \rightarrow \left\vert g\right\rangle $
occur for several times. In addition, we show one of the applications of the
kind of W class states, teleporting an entangled state of atomic ensembles
with unknown coefficients to many distant parties, either one of which
equally likely receives the transmitted state.

Let us have a look at the generalized form $\left| W_M\right\rangle $ of the
W class state in multi-qubit systems. In Ref. \cite{Dur}, the state is
defined as 
\begin{equation}
\left| W_M\right\rangle =\left( 1/\sqrt{M}\right) \left| M-1,1\right\rangle ,
\eqnum{1}
\end{equation}
where $\left| M-1,1\right\rangle $ denotes the totally symmetric state
including $M-1$ zeros and $1$ one. For example, we obtain $M=3$, 
\begin{equation}
\left| W_3\right\rangle =\left( 1/\sqrt{3}\right) \left( \left|
001\right\rangle +\left| 010\right\rangle +\left| 100\right\rangle \right) .
\eqnum{2}
\end{equation}

The basic element of this scheme is an ensemble of many identical alkali
atoms with a Raman type $\Lambda $-level configuration coulped by a pair of
optical fields with the Rabi frequencies $\Omega $ and $\omega $,
respectively, shown as Fig. 1, the experimental realization of which can be
either a room-temperature dilute atomic gas \cite{Jul,Phi} and a sample of
cold trapped atoms \cite{Liu,Roch}. We continue to use the symbols and
corresponding definitions shown in Refs. \cite{Duan2,Duan3}. A pair of
metastable lower states $\left\vert g\right\rangle $ and $\left\vert
s\right\rangle $ can be achieved, for example, in hyperfine or Zeeman
sublevels of electronic ground states of alkali atoms. The atoms in the
ensembles are initially prepared to the ground state $\left\vert
g\right\rangle $ through optical pumping. The transition $\left\vert
g\right\rangle \rightarrow \left\vert e\right\rangle $ is coupled by the
classical laser with the Rabi frequency $\Omega $ and the forward scattering
Stokes light comes from the transition $\left\vert e\right\rangle
\rightarrow \left\vert s\right\rangle $ \cite{Duan2}. The pumping laser is
shined on all atoms so that each atom has an equal small probability to be
excited into the state $\left\vert s\right\rangle $ through the Raman
transition. After the atomic gas interacts with a weak pumping laser, there
will be a special atomic mode $s$ called the symmetric collective atomic mode%
\begin{equation}
s=\left( 1/\sqrt{N_{a}}\right) \sum_{i=1}^{N_{a}}\left\vert g\right\rangle
_{i}\left\langle s\right\vert ,  \eqnum{3}
\end{equation}%
where $N_{a}\gg 1$ is the total atom number. In particular, an emission of
the single Stokes photon in a forward direction results in the state of
atomic ensembles given by $s^{+}\left\vert vac\right\rangle $, where the
ensemble ground state $\left\vert vac\right\rangle =\otimes _{i}\left\vert
g\right\rangle _{i}$. The scheme for preparation of W class of maximally
entangled states between atomic ensembles works in the following way (seeing
Fig 2):

1 The first step is to share an EPR type of entangled state between two
distant ensembles $1$ and $2$ using the scheme shown in Ref. \cite{Duan2}.
The ensembles are illuminated by a weak pumping laser pulse which couples
resonantly the transition $\left\vert g\right\rangle \rightarrow \left\vert
e\right\rangle $ and we look at the spontaneous emission light from the
transition $\left\vert e\right\rangle \rightarrow \left\vert s\right\rangle $%
, whose frequency is assumed to be different to the pumping laser. There are
two pulses with the frequencies $\omega _{pump}$ and $\omega _{repump}$,
respectively, which correspond to the pumping and repumping process. Here
two pumping laser pulses excite both ensembles simultaneously and with
probability $p_{c}$ the projected state of the ensembles $1$ and $2$ is an
EPR state with the form 
\begin{equation}
\left\vert \psi \right\rangle _{12}=\left( s_{1}^{+}+e^{i\phi
_{12}}s_{2}^{+}\right) /\sqrt{2}\left\vert vac\right\rangle _{12},  \eqnum{4}
\end{equation}%
where $\phi _{12}=\phi _{2}-\phi _{1}$ is a difference of the phase shift
which is fixed by the optical channel connecting the two ensembles, and $%
\left\vert vac\right\rangle _{12}$ denotes that both ensembles are in the
ground state $\left\vert g\right\rangle $.

2 We then connect the other two distant ensembles $2$ and $3$. Since the
ensemble $3$ is prepared to the ground state $\left\vert g\right\rangle $,
the whole system is described by the state $\left\vert \psi \right\rangle
_{12}\otimes \left\vert vac\right\rangle _{3}$. Here two pumping pulses
excite both ensembles simultaneously and the forward scattering Stokes light
from both ensembles is combined at the $50/50$ beam splitter (BS) after some
filters which filter out the pumping laser pulses with the outputs detected
by the two single-photon detectors D1 and D2, respectively. If one photon is
detected by either of the detectors, we obtain the state 
\begin{equation}
\left\vert \psi \right\rangle _{123}=\left( s_{2}^{+}+e^{i\phi
_{23}}s_{3}^{+}\right) /\sqrt{2}\left\vert \psi \right\rangle _{12}\otimes
\left\vert vac\right\rangle _{3}.  \eqnum{5}
\end{equation}%
Otherwise, we need to manipulate repumping pulses to the transition $%
\left\vert s\right\rangle \rightarrow \left\vert e\right\rangle $ on the
three ensembles and set them back to the ground state. Then, we repeat the
steps 1 and 2 until finally we obtain a click in either of the two detectors.

3 A repumping laser pulse with the frequency $\omega _{repump}$ is applied
to ensemble $2$. If one excitation is registered from it, we succeed and go
on with the next step. Otherwise, we need to repeat the above steps until we
get the three ensembles in the entangled state $\left\vert W^{\prime
}\right\rangle _{123}$ successfully. 
\begin{eqnarray}
\left\vert W^{\prime }\right\rangle _{123} &=&s_{2}\left( s_{2}^{+}+e^{i\phi
_{23}}s_{3}^{+}\right) \left( s_{1}^{+}+e^{i\phi _{12}}s_{2}^{+}\right)
\left\vert vac\right\rangle _{123}  \eqnum{6} \\
&=&\left( s_{1}^{+}+2e^{i\phi _{12}}s_{2}^{+}+e^{i\phi
_{13}}s_{3}^{+}\right) \left\vert vac\right\rangle _{123},  \nonumber
\end{eqnarray}%
where $s_{2}s_{2}^{+}s_{2}^{+}\left\vert vac\right\rangle _{2}=2\frac{N_{a}-1%
}{N_{a}}s_{2}^{+}\left\vert vac\right\rangle _{2}\simeq 2s_{2}^{+}\left\vert
vac\right\rangle _{2}$ ($N_{a}\gg 1$).

4 However, it is evident that the state $\left| W^{\prime }\right\rangle
_{123}$ above does not belong to the W class of maximally entangled states
shown in Eqs. (1) and (2). Then we connect the ensembles $1$ and $3$ using
the same way in the step 2, and apply a repumping laser pulse to the
ensemble $1$ after a click in D4 or D5. If there is one excitation is
registered by D6, we obtain the W class of maximally entangled state 
\begin{eqnarray}
\left| W\right\rangle _{123} &=&s_1\left( s_1^{+}+e^{i\phi
_{13}}s_3^{+}\right) /2\sqrt{3}\left| W^{\prime }\right\rangle _{123} 
\eqnum{7} \\
&=&\left( s_1^{+}+e^{i\phi _{12}}s_2^{+}+e^{i\phi _{13}}s_3^{+}\right) /%
\sqrt{3}\left| vac\right\rangle _{123}.  \nonumber
\end{eqnarray}

5 Similarly, suppose that the ensembles $1$ and $2$ are in the EPR\ state $%
\left\vert \psi \right\rangle _{12}$, to entangle $n$ ensembles in the W
class state, firstly we connect the ensembles $i$ and $i+1$, and then repump
the ensemble $i$ ($i$ from $2$ to $n-1$) after a right click orderly. It
needs to repeat the steps 2 and 3 for $n-2$ times to obtain the $n$-party W
class of non-maximally entangled state 
\begin{eqnarray}
\left\vert W^{\prime }\right\rangle _{1...n} &=&\prod_{i=2}^{n-1}s_{i}\left(
s_{i}^{+}+e^{i\phi _{i,i+1}}s_{i+1}^{+}\right) \left( s_{1}^{+}+e^{i\phi
_{12}}s_{2}^{+}\right) \left\vert vac\right\rangle _{1...n}  \nonumber \\
&=&\left( s_{1}^{+}+2\sum_{i=2}^{n-1}e^{i\phi _{1i}}s_{i}^{+}+e^{i\phi
_{1n}}s_{n}^{+}\right) \left\vert vac\right\rangle _{1...n}.  \eqnum{8}
\end{eqnarray}%
The difference of phase shift $\phi _{1i}$ is fixed by the possible
asymmetry of the setup and in principle can be measured. So we can put some
suitable phase shifters with relative phase shift to counteract it. Then we
need to repeat the manipulation above to the ensembles $1$ and $n$. Thus, we
can entangle $n$ ensembles in the W class of maximally entangled states 
\begin{eqnarray}
\left\vert W\right\rangle _{1...n} &=&\frac{1}{2\sqrt{n}}s_{1}\left(
s_{1}^{+}+e^{i\phi _{1n}}s_{n}^{+}\right) \left\vert W^{\prime
}\right\rangle _{1...n}  \eqnum{9} \\
&=&\frac{1}{\sqrt{n}}\sum_{i=1}^{n}e^{i\phi _{1i}}s_{i}^{+}\left\vert
vac\right\rangle _{1...n}.  \nonumber
\end{eqnarray}%
Note $\left\vert W^{\prime }\right\rangle $ in Eqs. (6) and (8) are not
normalized. The normalization constant for Eq. (8) is $1/\sqrt{4n-6}$.

Now, we consider the efficiency of this scheme, which is usually described
by the total generation time. Since the probability for getting a click of
either of two detectors is given by $p_c$, we entangle $n$ ensembles in the
W class state with the probability $\left( p_c\right) ^n$. In the generation
process, the dominant noise is the photon loss, which includes the
contributions from the channel attenuation, the spontaneous emissions in the
atomic ensembles, the coupling inefficiency of Stokes light into and out of
the channel, and the inefficiency of the single-photon detectors which can
no perfectly distinguish between one and two photons. All the above noise is
described by an overall loss probability $\eta $. Due to the noise, the
total generation time is represented by $T\sim t_0/\left[ \left( 1-\eta
\right) ^{2n-1}p_c^n\right] $, where $t_0$ is the light-atom interaction
time. And the generation time increases with the number of ensembles
exponentially by the factor $1/\left[ \left( 1-\eta \right) ^2p_c\right] $.

Also with the noise , the state of the ensembles is actually described by 
\begin{equation}
\rho _n=\frac 1{c_n+1}\left( c_n\rho _{vac}+\left| W\right\rangle
_{1...n}\left\langle W\right| \right) ,  \eqnum{10}
\end{equation}
where the vacuum coefficient $c_n$ is basically given by the conditional
probability for the inherent mode-mismatching noise contribution (please
seeing Ref. \cite{Duan4} for details) and $\rho _{vac}$ stands for the
vacuum component with no excitation in the ensembles $n-1$ and $n$.

Now we would like to use this W class state in one of the communication
protocols. Imagine that we need to spread an entangled state between atomic
ensembles with unknown coefficients to more than one parties. We choose a
three-party protocol by way of example and it will become evident that there
are many users that will work equally well.

Suppose there are three parties, the sender Alice, the receivers Bob and
Carol. Alice entangles the ensembles $1$, $2$ and $3$ ($4$, $5$ and $6$) in
the W class state $\left| W\right\rangle _{123}$ $\left( \left|
W\right\rangle _{456}\right) $. The pair of ensembles $i$ and $i+3$ $(i$
from $1$ to $3$) are put in the same place so that the ensembles $1$, $2$, $%
3 $ and $4$, $5$, $6$ can be connected through the same optical channel,
which fixed the phase shifts to be the same. So the states $\left|
W\right\rangle _{123}$ and $\left| W\right\rangle _{456}$ can be shown using
Eq. (7), where $\phi _{13}=\phi _{46}$ and $\phi _{12}=\phi _{45}$. The
ensembles $2$ and $5 $ are sent to Bob, $3$ and $6$ to Carol, $1$ and $4$
are left for herself.

Alice wants to teleport an atomic \textquotedblleft
polarization\textquotedblright\ state $\left\vert \varphi \right\rangle
_{un}=\left( \alpha s_{L}^{+}+\beta s_{R}^{+}\right) \left\vert
vac\right\rangle $ \cite{Duan2}, with unknown coefficients $\alpha $ and $%
\beta $, $\left\vert \alpha \right\vert ^{2}+\left\vert \beta \right\vert
^{2}=1$. Then she connects the ensembles $L$\ and $1$, $R$ and $4$ by
manipulating the repumping lase pulses with frequency $\omega _{repump}$ on
them synchronistically (shown in Fig 3). If the ensemble is in the
metastable state after the repumping pulse, the transition $\left\vert
e\right\rangle \rightarrow \left\vert s\right\rangle $ will occur {\it %
determinately}. The forward-scattering Stokes pulses are interfered at the
beam splitters after the filters. If Alice get two clicks, one in D1 or D2,
and the other in D3 or D4, the process is finished and the state of the
ensembles of Bob and Carol is shown as 
\begin{equation}
\left[ e^{i\phi _{13}}\left( \alpha s_{3}^{+}+\beta s_{6}^{+}\right)
+e^{i\phi _{12}}\left( \alpha s_{2}^{+}+\beta s_{5}^{+}\right) \right]
\left\vert vac\right\rangle _{2356}.  \eqnum{11}
\end{equation}%
Otherwise, they should prepare the W class states and repeat the above steps
until there are two correct clicks. Thus the state is teleported to the two
receivers, either one of which equally likely receives the transmitted
state, and similar to the scheme shown in Ref. \cite{Duan2}, the
teleportation fidelity would be nearly perfect.

It may be worth mentioning that if Bob and Carol perform a measurement, then
one of them can recover the state with unit fidelity in a probabilistic
manner. For example, Carol measures her ensembles $3$ and $6$ using two
repumping laser pulses (seeing Fig 3 for details), and the Stokes pulses are
collected by the detectors D5 and D6. If she obtains the original state $%
\left| \varphi \right\rangle _{un}$, there will be one click in either of
the two detectors. Else, one excitation is registered from each ensembles,
the original state is obtained by the other receiver Bob.

Finally, we have a brief conclusion. In this report, we describe an
experimental scheme of entangling many atomic ensembles in the W class of
maximally entangled states through laser manipulation. This protocol fits
well the status of the current experimental technology. In addition, we show
one of the applications of the kind of W class states, teleporting an
entangled state between atomic ensembles with unknown coefficients to two
distant parties, either one of which equally likely receives the transmitted
state.

We thank L.-M. Duan for helpful discussion and Y.-S. Zhang and Z.-W. Zhou
for their stimulating comments on an earlier version. This work was funded
by National Fundamental Research Program (2001CB309300), National Natural
Science Foundation of China, the Innovation funds from Chinese Academy of
Sciences, and also by the outstanding Ph. D thesis award and the CAS's
talented scientist award entitled to Luming Duan.

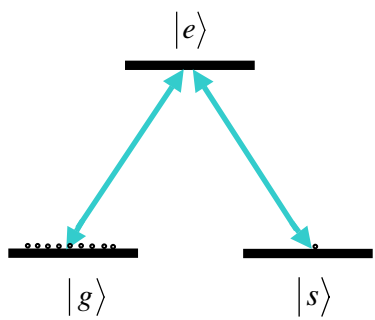
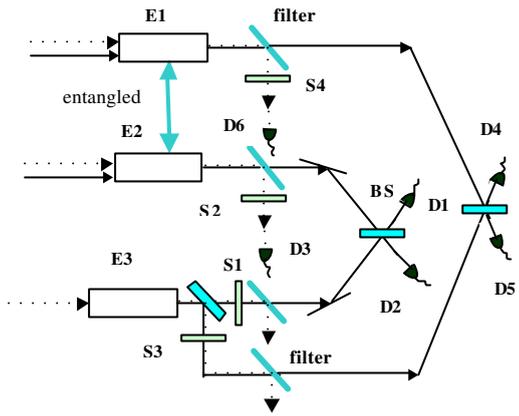
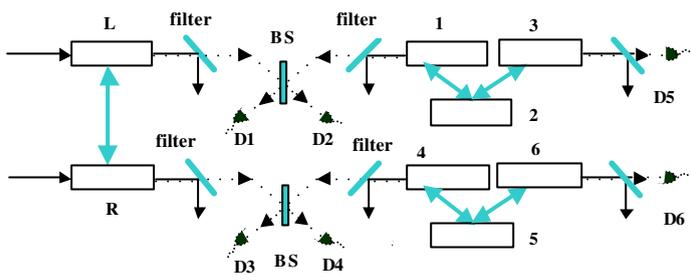
\end{document}